\documentclass[preprint,onecolumn,12pt,3p,authoryear]{elsarticle}

\usepackage{amssymb}
\usepackage{amsmath}
\usepackage{mathtools, cuted}
\usepackage[colorlinks=true,linkcolor=black, citecolor=blue, urlcolor=blue]{hyperref}
\usepackage{mathtools}

\newcommand{\ret}{\mathrm{Re}_{\tau}}

\DeclarePairedDelimiter{\norma}{\lVert}{\rVert}

\usepackage{lineno}

\journal{International Journal of Multiphase Flows}

\begin{document}

\begin{frontmatter}



\title{Investigating the magnitude and temporal localization of inertial particle mixing in turbulent channel flows}


\author[1]{Davide Perrone\corref{cor1}}\ead{davide.perrone@polito.it}
\author[2]{J.G.M. Kuerten}
\author[3]{Luca Ridolfi}
\author[1]{Stefania Scarsoglio}

\cortext[cor1]{Corresponding author}

\affiliation[1]{organization={Department of Mechanical and Aerospace Engineering, Politecnico di Torino},
            city={Turin},
            postcode={10129}, 
            country={Italy}}
            
\affiliation[2]{organization={Department of Mechanical Engineering, Eindhoven University of Technology},
			addressline={P.O. Box 513, 5600 MB}, 
            city={Eindhoven},
            country={The Netherlands}}

\affiliation[3]{organization={Department of Environmental, Land and Infrastructure Engineering, Politecnico di Torino},
            city={Turin},
            postcode={10129}, 
            country={Italy}}

\begin{abstract}
Mixing of inertial point particles in a turbulent channel flow at $\ret = 950$ is investigated by means of direct numerical simulations. We consider inertial particles, at varying Stokes number, released from pairs of sources located at different positions inside the channel and analyze the rate at which particles come into close proximity to each other. To do so, we employ a Lagrangian framework, which is suitable for the analysis of trajectories and in general for the study of mixing and dispersion problems. 
By varying the release position of particles along the wall-normal direction we obtain a thorough description of mixing in an anisotropic turbulent flow. Moreover, we analyze the effects of particle inertia and show that these are not univocal but also depend on the position and alignment of the sources, owing in particular to the dependence of the flow timescales on the distance from the wall.
\end{abstract}

%


\end{frontmatter}

\section{Introduction}
\label{sec:intro}

The mixing of inertial particles advected by fluid flows is strongly affected by turbulence, which enhances the intensity at which particles are displaced from their origin and driven towards each other (\cite{warhaft2000arfma,falkovich2001rmpa,dimotakis2005arfm}). In anisotropic wall-bounded flows, that are ubiquitous in nature and in industrial processes, the effects of turbulent fluctuations and of the mean flow intertwine and further complicate the analysis of mixing (\cite{nguyen2018aj}). Several numerical and experimental studies have been performed to investigate the mixing of scalar quantities and inertial particles, focusing on the analysis of the effects of particle properties, their release configuration and the features of the underlying flow (\cite{jayesh1992pffd,panchapakesan1993jfm,eswaran1988pf,yeung2002pf,eisma2021jfm}). The interplay of turbulence and particle inertia leads in general to strongly uneven concentration of inertial particles, which tend to accumulate in low vorticity zones of the flow domain \citep{ictr2008prla,mortimer,oujia2020jfm,brandt}, forming clusters that are relatively short-lived \citep{liu2020jfm}. 
Overall, the inertia of particles adds complexity to their behavior. The most relevant parameter in this regard is the ratio between the typical timescale of particles $\tau_p = \rho_p D_p^2/\left(18\nu\right)$ and the smallest local timescale of the flow $\tau_{\eta}$ (the Kolmogorov time), \textit{i.e.} the Stokes number $St_K$.
While at low Stokes numbers $St_K$ (much lower than unity) particles are akin to tracers, when $St_K$ becomes $\mathcal{O}(1)$ phenomena such as preferential concentration and turbophoresis are at a peak. Finally, at even higher Stokes number particles behave ballistically and react weakly to turbulence.
Such mechanisms may lead to local concentrations of particles that largely exceed the average, global, concentration and have been recognized as key players in cloud formation, environmental flows subjected to pollution, and combustion processes (\cite{crowe1988pieacs,chein1988aj,clouds1998,pan2013bm,lau_nathan_2014}).

Lagrangian statistics of non-tracer particles have been extensively studied, highlighting the impact of inertia on the magnitude of the velocity fluctuations of particles \citep{marchioli08}. Particles with higher inertia react weakly to small-scale velocity fluctuations of the carrier fluid. Furthermore, their velocity distribution is skewed, resulting in a net motion towards the wall, usually termed \textit{turbophoresis}. 
\cite{marchioli02} analyzed how sweeps and ejections play a crucial role in determining turbophoresis: sweeps transfer particles into the near-wall region and ejections move them back into the outer flow, but may fail to do so as inertial particles become concentrated into very elongated streamwise-aligned streaks associated with negative streamwise velocity fluctuations (as also observed experimentally by \cite{fong19}). Furthermore, sweeps and ejections must retain enough spatial and temporal coherence in order to actively displace inertial particles.

Besides inertia effects on the interaction of particles with the near wall velocity field, another important feature of non-tracer particles is their tendency to preferentially sample only certain regions of the domain and thus locally increase or reduce the concentration, which is named inertial clustering. The latter turns out to be strongly influenced by the interplay between particle response time $\tau_p$ and the timescale of the relevant eddies of the flow. Clustering has been quantified using several methods, including box-counting methods \citep{rouson01}, Voronoi tessellation methods \citep{monchaux10,monchaux12,liu2020jfm} and wavelet filters \citep{bassenne17}. 
Numerical simulations and experiments, along with analytical approaches \citep{esmaily16}, confirm that the most intense degree of clustering is attained when the particle timescale matches that of the smallest scale of the fluid (that is, $St_K = \mathcal{O}(1)$). Particles heavier than the carrier fluid are swept away from vortex cores.
\cite{oka21} demonstrated, for light particles which instead accumulate around vortex cores, how the multiscale nature of homogeneous turbulence affects clustering. In particular, particles with different response times $\tau_p$ cluster around vortex cores of different sizes.
In non-homogeneous turbulence the flow properties are not constant in space, but depend on position. Consequently, clustering is influenced by the instantaneous location of particles. In channel flow, both the local Kolmogorov timescale and the integral timescale of particle velocities have to be taken into account for an accurate description of particle behavior \citep{marchioli06}.

Together, timescale features influence dispersion and mixing of inertial particles. \cite{bec10} investigated pair separation of light and heavy particles in homogeneous turbulence, finding that the inertia of particles is relevant on short timescales, while a tracer-like behavior is recovered for longer times. In channel flow, \cite{pitton12} found superdiffusive behavior due to mean shear for long times, a definite influence of the direction of the initial separation of particles and a non trivial dependence on wall-distance of the rate of separation of pairs.

In this work, we propose a Lagrangian approach to characterize the mixing of particles released from pairs of point-like sources in turbulent channel flow, in order to perform a thorough description of mixing properties with respect to different parameters (\textit{i.e.}, particle inertia, position and distance of the sources). In particular, we measure the number of particles, coming from distinct sources, that come at a mutual distance shorter than a threshold $R$, and are therefore able to define a probability $E$ that initially distant particles come in close proximity to each other. By doing so, we aim to quantify the influence of the different phenomena concerning inertial particles and their role in determining the evolution of mixing.

Lagrangian approaches appear suitable to further our comprehension of mixing processes and have widely been applied to the analysis of massless tracer particles \citep{raissi2019prf,llamas2020cit,liu2020jfm,perrone1,perrone2,schneide22}. 
Similar approaches to the one presented in this work have been employed by \cite{iacobello2019jfm}, to quantify mixing of passive tracers using a network-based perspective, and by \cite{rypina2017npga}, to provide a straightforward measure for the mixing potential in an oceanic flow. On the contrary, the characterization of the mixing properties of inertial particles in non-homogeneous, wall-bounded flows using a Lagrangian perspective has not yet been carried out to the best of our knowledge.

We employ direct numerical simulation (DNS) of turbulent channel flow at $\ret = 950$, together with Lagrangian tracking. We consider point particles with inertia, which do not affect the flow and do not undergo collisions. 
We also performed simulations at a lower Reynolds number $\ret = 590$ to show the variation of mixing properties with respect to this parameter. Direct numerical simulations have become an invaluable tool due to their accuracy, the insight on complex flow features and the relative ease with which several flow parameters can be varied.

We consider inertial particles with different diameters, resulting in four cases at distinct Stokes numbers ${St^+} = 0.2$, 1, 5, 25, where $St^+ = \tau_p^+ = St_K \tau_{\eta}^+$ is the Stokes number in wall units, which is used instead of $St_K$ as the latter is not constant across the channel height, the Kolmogorov time $\tau_p$ being $y$-dependent. Therefore we test a large range of different behaviors, from tracer-like particles to ones with large inertia. 
Furthermore, by varying the location of the two sources, we provide a thorough description of the influence of both the absolute position of the sources and of their mutual distance, both of which are important as the effect of the convection performed by the mean velocity profile and the turbulence structure on mixing is strongly anisotropic.

This work is organized as follows. Section \ref{sec:methods} describes the DNS setup, the release configuration of inertial particles and the method employed to compute the probability that particles released from distinct sources encounter. Section \ref{sec:results} reports the main results of our analysis. Section \ref{sec:concl} recaps the main findings and gives some final remarks. Finally, in \ref{app:a} we provide brief results at a different, lower Reynolds number. In \ref{app:c} we report Eulerian statistics of channel flow, with the comparison with previous simulations used to validate our code. Finally, in \ref{app:b} we analyze the sensitivity of our approach with respect to the threshold range of the interaction $R$.

\section{Methods}
\label{sec:methods}

\subsection{DNS and particle tracking}
In order to measure the mixing of Lagrangian point particles released in a non-homogeneous flow, we performed a pseudo-spectral numerical simulation of a fully developed turbulent channel flow and tracked inertial point particles released inside it.
The Navier-Stokes equations in rotation form
\begin{equation}
\label{eq:ns1}
\nabla\cdot\mathbf{u} = 0,
\end{equation}
\begin{equation}
\label{eq:ns2}
\frac{\partial \mathbf{u}}{\partial t} + \frac{1}{\rho}\nabla P = \mathbf{f} - \boldsymbol\omega\times \mathbf{u} + \nu \nabla^2\mathbf{u},
\end{equation}
where $\mathbf{u}$ is the velocity, $\rho$ the fluid mass density, $\boldsymbol\omega = \nabla \times \mathbf{u}$ the vorticity, $\nu$ the kinematic viscosity and $P$ the total pressure, were solved in a box of size $2\pi\delta \times 2\delta \times \pi\delta$, with $\delta$ being the channel half-height. 
Periodic boundary conditions were set along the streamwise $x$ and spanwise $z$ directions, while the no-slip condition was imposed along the wall-normal $y$ direction at the two walls $y=0$ and $y = 2\delta$. We performed the direct numerical simulation at a Reynolds number $\ret = 950$, where $\ret = \delta u_{\tau}/\nu$ is the Reynolds number based on the friction velocity $u_{\tau} = \sqrt{\tau_w/\rho}$ ($\tau_w$ is the wall shear stress). Details regarding the effects of the Reynolds number are given in \ref{app:a}, with results of simulations at $\ret = 590$.
Equations \eqref{eq:ns1}-\eqref{eq:ns2} were discretized in space using Fourier base functions in the periodic $x$ and $z$ directions and a Chebyshev-$\tau$ approach in the wall-normal one. The number of Fourier polynomials is 768 in both homogeneous directions, while 384 Chebyshev polynomials were used to discretize quantities along the wall-normal direction. Nonlinear terms were computed in physical space and dealiased using the $3/2$ rule and were explicitly advanced in time with a second order Runge-Kutta method, while linear terms were integrated implicitly with a Crank-Nicholson scheme. The time-step of the simulation was $\Delta t^+ = 0.095$. A sketch of the channel geometry is given in figure \ref{fig:1}(a).
The numerical scheme adopted is the same as defined in \cite{kuerten13} and the simulation was run until a state of fully developed turbulence was reached and flow quantities reach statistical stationarity. Average flow quantities were compared with those of previous simulations. In particular, average and root mean square velocity components were found in very good agreement with those of \cite{hoyas08}, although the domain is smaller while the streamwise and spanwise resolutions are higher. Comparison between statistical quantities of this simulations and those used as reference can be found in \ref{app:c}.

In order to obtain sets of trajectories $\mathbf{x}(\mathbf{x}_0, t)$ we integrated the velocity of each particle $\mathbf{v}(\mathbf{x}_0, t)$ with the same explicit Runge-Kutta scheme as used for the nonlinear terms of the Navies-Stokes equations, starting from the release location $\mathbf{x}_0$. Inertial particles, that are treated as point particles in the simulations, do not match the fluid velocity and are instead subject to a drag force. Accordingly, for particles with non-negligible particle Reynolds number $\mathrm{Re}_p$, the equation for the particle velocity $\mathbf{v}$ is
\begin{equation}
\label{eq:drag}
\frac{\mathrm{d}\mathbf{v}}{\mathrm{d}t} = \frac{\mathbf{u}(\mathbf{x}(\mathbf{x}_0, t), t) - \mathbf{v}(\mathbf{x}_0, t)}{\tau_p}\left(1 + 0.15\,\mathrm{Re}_p^{0.687}\right),
\end{equation}
where $\mathrm{Re}_p = d_p\left\lvert\lvert\mathbf{u} - \mathbf{v}\right\rvert\rvert/\nu$, $\tau_p = \rho_p d^2_p/\left(18\rho\nu\right)$, and $d_p$ and $\rho_p$ are the particle diameter and mass density, respectively \citep{maxeyriley,kuerten06}. The ratio between the particle relaxation time $\tau_p$ and the characteristic time scale of the flow is the Stokes number $\mathrm{St} = \tau_p^+$, where the $^+$ superscript indicates normalization with wall units. The density ratio $\rho_p/\rho$ between the particles and the carrier fluid is 769.23; the Stokes number is varied by changing the diameter of particles, thus resulting in diameters ranging from $d^+_p = 0.068$ at $St^+ = 0.2$ to $d^+_p = 0.764$ at $St^+ = 25$. Collisions with the wall are treated elastically. Gravity effects were not included in order to focus only on turbulence-induced features of particles dynamics, which are easily overwhelmed by settling effects if gravity is included.

\subsection{Measure of particle encounters}

\begin{figure}
\centering
\includegraphics[width = .85\textwidth]{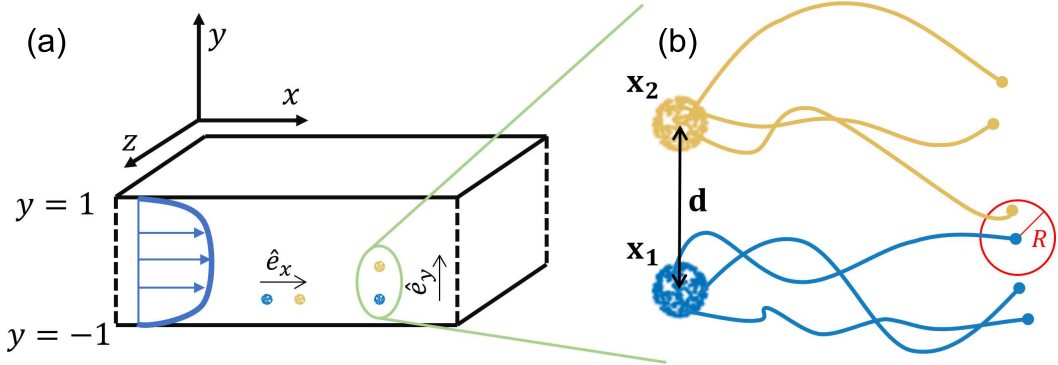}
\caption{a) Channel geometry and coordinate system; two pairs of sources are represented, one aligned along the streamwise direction, the other along the wall-normal one. b) Schematic depiction of how encounters are determined starting from particle trajectories. \label{fig:1}}
\end{figure} 

We aim to measure the rate at which clouds of tracers, released from different sources, encounter and mix together. We released $N_p = 256$ tracers from spherical, point-like sources whose size is finite and of the order of the smallest scales of the flow. We set the radius of the spherical sources equal to 1 in wall units, which is comparable to the Kolmogorov length scale (although the latter is not constant across the channel heigth). In this way, the evolution of trajectories reflects the properties of a local region while avoiding that particle velocities remain correlated for very long times, as would instead happen if all particles were released in the same point. This is done solely to accelerate the computation of statistics by releasing a relatively large set of particles from each source and measuring the interaction as will be detailed in the following. The aim of this paper is to investigate the mixing behavior of particles at particle volume fractions that are so low that their effect on the fluid and their interaction are negligible. In the simulations presented in the paper the number of particles is only high to calculate the mixing of all particle pairs in one simulation and to increase the statistical accuracy of the results.

To quantify mixing between two distinct sources, we measured the number of particle pairs coming from the two plumes that are closer than a threshold distance $R$, as shown in figure \ref{fig:1}(b). We defined this number, normalized by its maximum attainable value $N_p^2$, as the encounter probability $E$. An encounter probability of $1$ would be obtained only in the presence of perfect mixing between two sources, \textit{i.e.} if all particles released from both sources become located in a small (of the order of the threshold $R$) region of space. 
For two plumes released from two distinct locations $\mathbf{x}_1$ and $\mathbf{x}_2$, the (time-dependent) encounter probability can be computed as
\begin{equation}
\label{eq:encprob}
E_{\mathbf{x}_1, \mathbf{x}_2}(t) = \frac{\lvert \left\lbrace \norma{\mathbf{x}_i(t) - \mathbf{x}_j(t)}\leqslant R : \mathbf{x}_i(0) = \mathbf{x}_1 \land \mathbf{x}_j(0) =\mathbf{x}_2\right\rbrace\rvert}{N_p^2}.
\end{equation}
$\mathbf{x}_{i/j}(t)$ is the trajectory of the $i/j$-th particle of the plume starting from $\mathbf{x}_{1/2}$ and $\norma{\cdot}$ is the Euclidean norm. From the definition it follows that $E_{\mathbf{x}_1, \mathbf{x}_2}(t) = E_{\mathbf{x}_2, \mathbf{x}_1}(t)$.

The choice of the threshold distance $R$ is nontrivial. The threshold should be sufficiently small, such that the clusters formed by the combined action of turbulence and particle inertia are accurately described and their effects are fully accounted for when measuring mixing. Since inertial clustering happens on relatively small scales (\textit{e.g.} tens of wall units), as it is the result of the action of small-scale features of the flow \citep{brandt}, we chose a threshold size $R^+ = 2$; an analysis of the sensitivity of our results with respect to the value of $R$ is reported in \ref{app:b}.

As the position of the two sources has the greatest influence on the evolution of the encounter probability, we thoroughly vary $\mathbf{x}_1$ and $\mathbf{x}_2$ across the channel. It is convenient to report the absolute position of one of the two sources, \textit{e.g.} $\mathbf{x}_1$, and the distance between the two sources $\mathbf{d} = \mathbf{x}_2 - \mathbf{x}_1$. As channel properties are statistically homogeneous along the $x$ and $z$ directions, only the wall-normal component of the position $y_1 = \mathbf{x}_1 \cdot \mathbf{\hat{e}}_y$ will be considered in the following. Instead, $\mathbf{d}$ is in general a vector quantity, since channel flow is anisotropic and mixing properties may depend on the relative orientation of the two sources. Meaningful values of $d = \lvert\lvert\mathbf{d}\rvert\rvert$ are restricted by the constraint that at least some particles come close enough that their mutual distance is less than $R$. Accordingly, the admissible values of $d$ are bound to that of $R$, since if $d \gg R$ the encounter probability becomes negligible. In this work, we analyze mixing with a distance between sources comprised between $d^+ = 4$ and $d^+ = 32$ (at which the encounter probability is already vanishing).

Finally, as particles are released into fully developed turbulent flow, their initial velocities are also subject to turbulent fluctuations and thus have locally different initial encounter properties. In order to achieve statistical convergence, for each set of parameters ($y_1$, $\mathbf{d}$, $St$) we analyze multiple pairs of sources, positioned at different $x$ and $z$ coordinates and at different times, with such spacing that their evolution is uncorrelated. 
Since channel properties are statistically homogeneous along the streamwise and spanwise directions and are statistically steady through time, the results can be averaged to obtain a statistically significant value of the encounter properties. For each set of parameters we analyze $N_r = 384$ independent pair of sources; accordingly, the total number of particles considered for each set of parameters is $2 N_p N_r \approx 2\cdot 10^5$ and the total number of particles simulated is approximately $7.68\cdot 10^7$ considering all Stokes numbers.

\section{Results}
\label{sec:results}

\begin{figure}
\centering
\includegraphics[width = \textwidth]{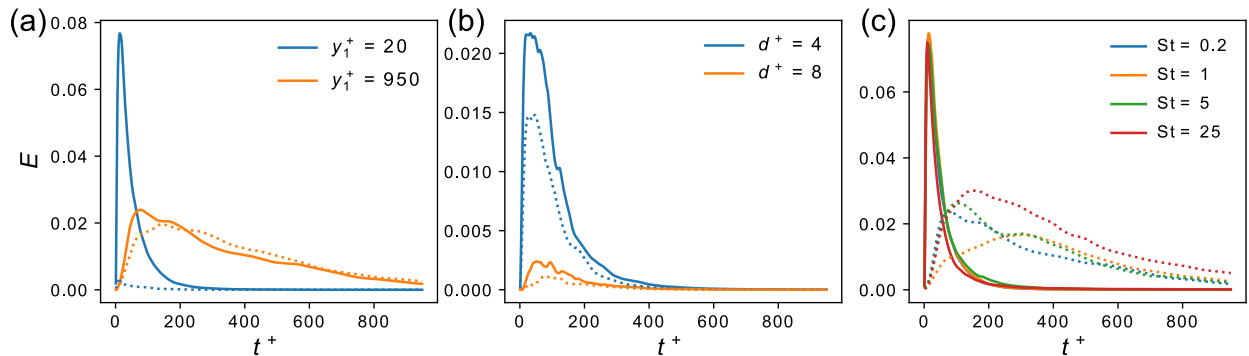}
\caption{a) Evolution of the encounter probability for pairs of sources at two different wall-normal locations, fixed $d^+ = 4$ and $St^+ = 0.2$ (solid lines indicate mixing of streamwise-aligned sources, dashed lines indicate mixing of sources aligned along $y$); b) $E(t)$ for source pairs at two different distances, $y_1^+ = 100$, $St^+ = 0.2$ (solid lines indicate mixing of streamwise-aligned sources, dashed lines indicate mixing of sources aligned along $y$); c) $E(t)$ for pairs of sources at fixed $\mathbf{d} = d\mathbf{\hat{e}}_x$, $d^+ = 4$ and varying Stokes number (solid lines are for source pairs at $y_1^+ = 20$, dashed lines at $y_1^+ = 950$.\label{fig:2}}
\end{figure}

The encounter probability of particles released from each pair of sources transiently grows owing to the action of turbulent diffusion and local shear. Indeed, turbulent velocity fluctuations and the local shear have both the effect (on average) to increase the size of the puffs of particles coming from each source over time. Still, locally and on short timescales, otherwise distant groups of particles are brought together by the local action of turbulence. After a peak of $E$ has been reached, particles are continuously driven apart because of turbulent diffusion and the mean shear due to the streamwise velocity profile, resulting in a vanishing encounter probability at long times. 

Figure \ref{fig:2} shows the typical evolution of the encounter probability for some pairs of sources, that are releasing particles whose Stokes number is 0.2 (panels (a) and (b)) and at varying Stokes number (panel (c)). In particular, we chose source pairs at two different wall-normal coordinates $y_1^+ = 20$ and $y_1^+ = 950$ in order to highlight the different properties of mixing found near the wall and at the center of the channel. Figure \ref{fig:2}(a) shows the encounter probability for two cases. In the first case the distance between sources is $\mathbf{d} = d\mathbf{\hat{e}}_x$, \textit{i.e.} the pair of sources is aligned along the streamwise direction $x$, while in the second case the distance is $\mathbf{d} = d\mathbf{\hat{e}}_y$ and as such sources are aligned along $y$; in both cases $d^+ = 4$. As noted before, channel properties are in general anisotropic (and especially near the wall) and also the properties of mixing depend on the direction of alignment of the sources. As can be seen in the figure, the encounter probability along the streamwise direction is higher near the wall, while it decreases away from it. 
Mixing of $y$-aligned sources ($\mathbf{d} = d\mathbf{\hat{e}}_y$) is negligible near the wall (where the differences of the local mean velocity are high) and increases towards the center. Most notably, encounter probabilities are independent of the direction near the center of the channel, where indeed channel properties tend towards homogeneity and the mean shear is close to zero. On the other hand, towards the channel walls mean shear is one of the main factors causing strong mixing of streamwise-aligned sources (especially due to their finite size) and weak mixing of $y$-aligned sources. 
In order to isolate the influence of mean shear from those of turbulent fluctuations, we also integrated the trajectories using only the mean turbulent velocity profile $U(y)$, effectively computing encounter properties that are determined solely by mean shear. Trivially, the encounter probability for $y$-aligned sources remains zero at all times (as particles travel along straight lines and do not come closer to each other). Instead, in the case of streamwise-aligned sources, a peak encounter value approximately equal to $E_{peak} = 0.2$ (with $d^+ = 4$, as is for figure \ref{fig:2}(a)) is reached independently of the $y$ coordinate of release except for the exact center of the channel where there is no shear. Turbulent velocity fluctuations cause the dispersion of particles, thus overall reducing the encounter rate, although this is also dependent on the intensity of mean shear and the size of sources (in the limit of very small source size, the effect of shear on the encounter rate becomes negligible).

Figure \ref{fig:2}(b) shows the effect of the value of $d$ on the encounter probability, at a single $y_1$ value. Both in the case of streamwise and wall-normal mixing, increasing the distance between sources greatly reduces the encounter probability.
Figure \ref{fig:2}(c) shows instead the streamwise ($\mathbf{d} = d\mathbf{\hat{e}}_x$, $d^+ = 4$) encounter probability at different Stokes numbers. While near the wall the encounter probability evolves almost independently of the Stokes number, near the center of the channel (dotted lines) a larger variability is seen with respect to the inertia of particles. Different Stokes numbers lead to different behavior with respect to inertial clustering, thus causing differences in the encounter rate. Near the wall (see solid lines) particle inertia does not seem to influence the encounter rate, possibly because encounters happen on a timescale so short that clustering still cannot influence mixing. A more detailed comparison between the Lagrangian timescales of particles and those of mixing will be performed below. Moreover, particles released near the wall tend to organize in elongated, streamwise-aligned streaks, that exert a predominant influence on mixing.

On the other hand, near the center of the channel (dotted lines) the encounter probability appears to be stronger at $St^+ = 25$ and lower at $St^+ = 1$, with $St^+ = 0.2$ and $St^+ = 5$ having an intermediate behavior. A different behavior is found in the case of $y$-aligned sources (not shown in the figure), where the encounter probability at $St^+ = 5$ reaches a lower peak than those at the other Stokes numbers. It appears that near the center of the channel the differences in particle inertia lead to non trivial behaviors with regard to mixing and the encounter probability. Inertial particles organize into small clusters also near the center of the channel, albeit not in elongated streaks \citep{hans2016ijmf}. 

The evolution of the encounter probability is the result of the action of competing phenomena, which initially bring a fraction of the particles together while at the same time dispersing them across larger and larger regions of the flow domain. 
Indeed, the combined and cumulative action of the different flow scales sampled by clouds of particles may, especially at short times and lengths, result in behaviors opposite to that of long term dispersion. As such, the short-term action of turbulence causes the interaction of particles released by the two sources, which determines an increase of the encounter probability $E$ due to the presence of converging motions inside the flow that drive particles together \citep{warhaft2000arfma}.
The peak with respect to time of the encounter probability is the point in the evolution of the two puffs of particles where these competing phenomena are in balance. Higher peak values $E_{peak}$ indicate that the local action of the flow is able to overcome, at least for a finite amount of time, the long time effects of dispersion which ultimately brings the encounter probability to zero. Similarly, the time $t_{peak}$ at which the peak happens is tightly linked to the timescales at which mixing happens.
In the following, we will analyze the properties of the peak of the encounter rate with respect to the position of the sources inside the channel and the particle properties.

\begin{figure}
\centering
\includegraphics[width = \textwidth]{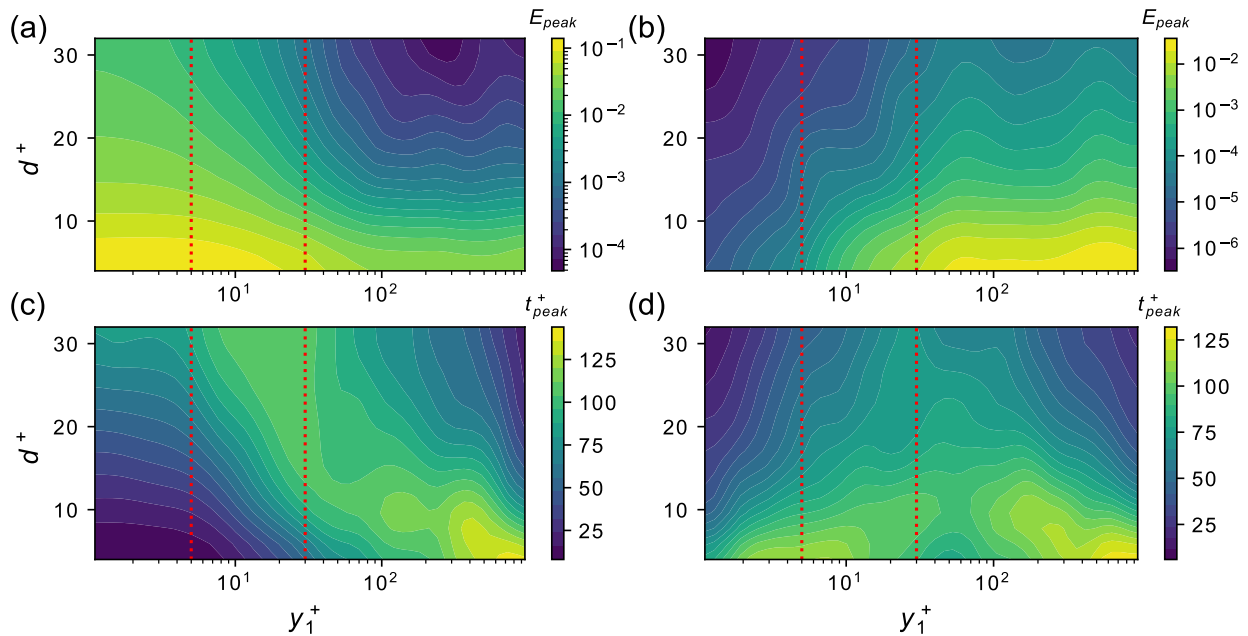}
\caption{Peak encounter probability for streamwise (a) and wall-normal (b) aligned sources as a function of $y_1^+$ and $d^+$ at $St^+ = 0.2$; the red lines mark the end of the viscous sublayer at $y^+ = 5$ and of the buffer layer at $t^+ = 30$. Peak time for streamwise (c) and wall-normal (d) aligned sources as a function of $y_1^+$ and $d^+$ at $St^+ = 0.2$. \label{fig:3}}
\end{figure}

Figure \ref{fig:3} explicitly highlights how the encounter probability rate (and in turn the peak time) is correlated to both $y_1^+$ (the distance from the wall) and $d^+$ (the reciprocal source distance), at a fixed $St^+ = 0.2$. In particular, \ref{fig:3}(a) and \ref{fig:3}(b) show streamwise (along $\mathbf{\hat{e}}_x$) and wall-normal (along $\mathbf{\hat{e}}_y$) peak encounter probability, respectively. As previously noted, mixing of streamwise-aligned sources is stronger near the wall, while mixing of sources aligned along $y$ increases away from it. The region where mixing of streamwise-aligned sources is the highest is confined well inside the viscous sublayer, where the dispersing action of turbulence is reduced. Accordingly, very high values (almost 0.2) of the encounter probability are reached for nearby sources. Instead, the wall-normal encounter probability is weaker (the largest peak values are of the order of 0.02), but the region where mixing is the strongest occupies a large portion of the channel, from the upper limit of the logarithmic region up to the centerline. While for the mixing of sources aligned along $x$ only turbulent diffusion determines the decrease of the encounter probability, especially at short times, for source pairs aligned along $\mathbf{\hat{e}}_y$ also mean shear is present, thus lowering the values of $E$.
Furthermore, mean shear reaches a maximum near the walls; indeed, puffs of particles released at different $y$ coordinates have strongly different mean velocity and are rapidly driven apart, resulting in almost zero peak encounter probability. The flatness of the turbulent velocity profile has the effect that outside of the logarithmic region this factor is of reduced importance.
Furthermore, $E_{peak}$ decreases as $d$ increases for both $\mathbf{\hat{e}}_x$ and $\mathbf{\hat{e}}_y$: indeed, if the two sources are more distant, the dispersive action of turbulence makes sure that fewer particles come in close proximity. Indeed, even if turbulence enables the initial growth of the encounter probability (as particles are dispersed and a fraction of them is driven close together), velocity fluctuations also disperse particles in all directions, thus the distance between sources tends on average to decrease the encounter probability (although it is still possible that some distant particles come in close proximity due to the features of the local velocity field).

Figures \ref{fig:3}(c) and \ref{fig:3}(d) show instead the time $t_{peak}$ at which the peak encounter probability is achieved, for mixing of sources aligned along $x$ and $y$, respectively. In the case of streamwise-aligned sources it appears that, especially for nearby sources (small $d$), the peak time is higher where mixing is less intense and \textit{vice versa}. The high intensity mixing taking place near the wall is achieved very quickly by particles that become entrained together in the elongated structures typical of the near-wall region.
As the distance between sources increases, longer peak times are observed for pairs of sources located in the buffer layer, although results may be skewed by the small number of encounters taking place.

\noindent In the case of $y$-aligned sources on the other hand, the peak time is higher near the center of the channel, where also the highest values of $E_{peak}$ are found, opposed to what was found for streamwise-aligned sources; nonetheless, peak times for pairs of sources near the wall are not negligible, especially at higher Stokes numbers as will be detailed in the following. The typical timescales of particle motion are in general higher near the center of the channel, thus resulting in increased peak times for the mixing of both streamwise and wall-normal aligned sources. 

\begin{figure}
\centering
\includegraphics[width = \textwidth]{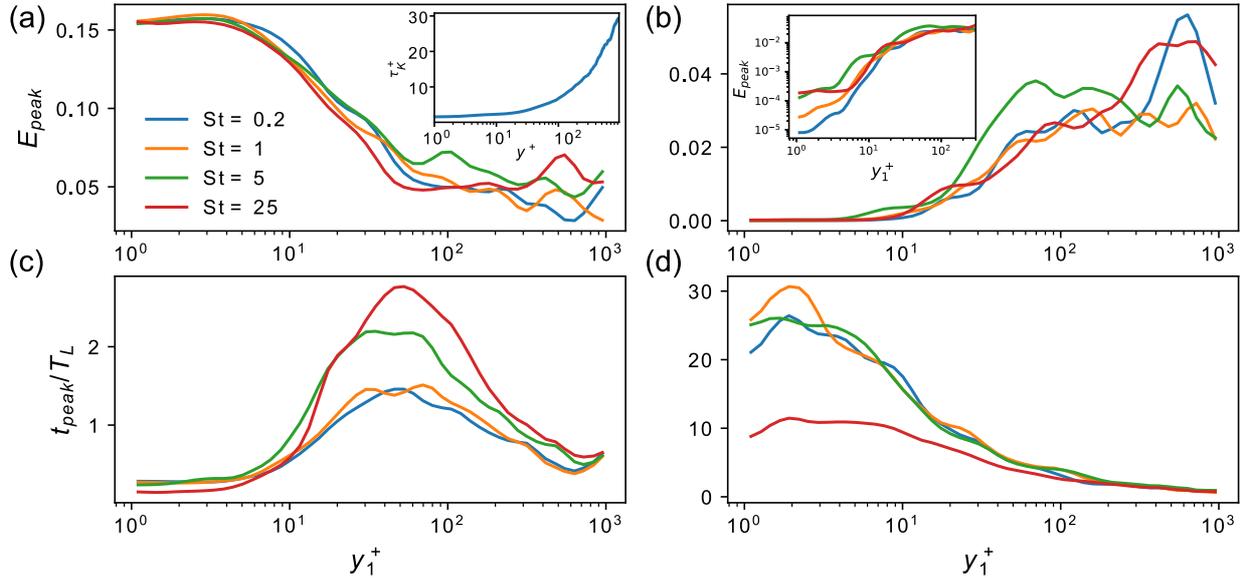}
\caption{Peak streamwise (a) and wall-normal (b) encounter probability $E_{peak}$ with $d^+ = 4$ at different Stokes numbers. The inset in the left panel shows the value of Kolmogorov time $\tau_p^+$ across the channel height, while the inset of the right panel shows a zoom of the data in figure \ref{fig:4}(b) with logarithmic vertical axis. Peak time of the streamwise (c) and wall-normal (d) encounter probability $t_{peak}$ with $d^+ = 4$ at different Stokes numbers. \label{fig:4}}
\end{figure}

As shown before in figure \ref{fig:2}(c), particle inertia has an effect on the encounter probability, as it affects clustering properties which determine the rate at which particles become confined to a smaller region of the fluid domain. Figure \ref{fig:4} shows a comparison of the peak encounter probability $E_{peak}$ and of its time $t_{peak}$ for different Stokes numbers (at a fixed distance $d^+ = 4$, which is also typical for the behavior at higher values of $d$). 

The peak encounter rate of source pairs located inside the viscous sublayer is independent of the Stokes number in the streamwise alignment case. On the other hand, farther away from the wall, streamwise-aligned sources (figure \ref{fig:4}(a)) have slightly stronger peaks at higher Stokes numbers. A noticeable increase of $E_{peak}$ can be found for particles with $St^+ = 5$ near $y^+ = 100$, while at near the center of the channel the same can be found for particles with $St^+ = 25$. Similar behavior has been found using different, uncorrelated particle simulations.

The wall-normal encounter probability (figure \ref{fig:4}(b)) has a slight dependence on the Stokes number at low $y^+_1$ values, where particles with higher Stokes numbers $St^+$ also have higher peak encounter rates. The near wall behavior is shown in the inset of figure \ref{fig:4}(b) using logarithmic scaling because of the very small values of $E$ attained by $y$-aligned sources. Instead, $E_{peak}$ is higher towards the center of the channel for particles with $St^+ = 0.2$ and $St^+ = 25$, but at slightly lower coordinates (again, around $y^+ = 100$) it is stronger at $St^+ = 5$.
The inset of figure \ref{fig:4}(a) shows the value of the Kolmogorov timescale $\tau_{\eta}^+$ in wall units across the channel height. Values of the timescale range from $\tau_{\eta}^+ \approx 5$ at around $y^+ = 100$ to $\tau_{\eta}^+ \approx 25$ at around $y^+ = 900$. Accordingly, the Stokes number scaled with Kolmogorov units for particles with $St^+ = 5$ located at around $y^+ = 100$ is approximately $St_K = St^+ / \tau_{\eta}^+ \approx 1$, while it is again $St_K \approx 1$ for particles with $St^+ = 25$ located near the center of the channel. 
Recalling that particles with values of $St_K$ of the order of unity tend to experience stronger clustering phenomena, it is very likely that the two slight increases of $E_{peak}$ found at $y^+ = 100$ for particles with $St^+ = 5$ and at $y^+ = 900$ for particles with $St^+ = 25$ are attributable to inertial clustering (see figures \ref{fig:4}(a)-(b)). 
This is only true away from the wall, whereas close to it the organization of particles into elongated structures aligned with the mean flow is the predominant effect leading to encounters (or the absence thereof). Indeed, no effect of particle inertia on the encounter properties is found for source pairs located close to the wall.
Enhanced encounter rates are also present near the center of the channel for particles with $St^+ = 0.2$, even if in this case particles are more akin to tracers given their very low inertia ($St_K\approx 10^{-2}$) and should not experience inertial clustering. We performed another set of particle simulations in order to confirm that this is not due to a too small sample size. Since in this region there is almost no mean shear and inertial clustering should not be a relevant factor for these smaller particles, these increased values of $E_{peak}$ may be caused by the action of the wall-normal velocity fluctuations, but the exact mechanisms behind this is not fully clear.

Figures \ref{fig:4}(c) and \ref{fig:4}(d) show the peak time $t_{peak}$ of the encounter rate for $x$- and $y$-aligned sources, respectively. Values are shown normalized by the local Lagrangian velocity timescale $T_L$, computed from the streamwise velocity in the case of $x$-aligned sources and from the wall-normal velocity in the case of $y$-aligned sources. The Lagrangian integral velocity timescale is computed as
\begin{equation}
\label{eq:tl}
T_{L,i}(y) = \int_0^{\infty} \rho(y,\tau) \mathrm{d}\tau =  \int_0^{\infty} \frac{\langle u_i(y,0) u_i(y,\tau) \rangle}{\sqrt{\langle u_i(y,0)^2\rangle\langle u_i(y,\tau)^2\rangle}} \mathrm{d}\tau,
\end{equation}
where the ensamble average $\langle \cdot \rangle$ is taken over particles released from the same $y$ coordinate, the $i$-th component of the velocity fluctuation $u_i$ is used and the integral is performed up to the first zero-crossing of the autocorrelation function $\rho$ because of the finiteness of the data \citep{stelz}.
First, encounters between streamwise-aligned sources positioned near the wall happen on a very short timescale, which may explain why the peak value $E_{peak}$ has little dependence on the Stokes number. Indeed, the timescale of encounters is so short that clustering phenomena are yet to arise in this case, while a stronger influence is made by the initial condition, as particle trajectories are still well into the ballistic regime (\textit{i.e.}, their velocity is strongly correlated with that at release) at the time of peak encounter.
As already stated, in the near-wall region the evolution of the encounter rate appears independent of the inertia of particles, as it takes place on shorter timescales and is rather an effect of the organization of particles into elongated streaks.
The peak time becomes longer than the Lagrangian timescale in the log layer, possibly due to the weaker mean shear which causes slower mixing of streamwise-aligned sources. Towards the center of the channel instead, peak encounters happen on a timescale shorter than the Lagrangian, even if shear is at a minimum there; accordingly, a more important role in this case is played by turbulent velocity fluctuations and inertial clustering effects. 
In the case of $y$-aligned sources instead, mixing near the wall happens on a timescale which is up to two orders of magnitude longer than for $x$-aligned sources. Indeed, here mean shear strongly inhibits encounters and only turbulent fluctuations may drive particles together (in very small quantities, as shown by the low values of $E_{peak}$ in figure \ref{fig:4}(b)). The peak time diminishes monotonically towards the center of the channel following the reduction of mean shear.

\section{Conclusions}
\label{sec:concl}
In this work we explored properties of mixing of inertial particles in turbulent channel flow. We defined a metric, the encounter probability $E$, which takes into account the spatial proximity of particles released by sources inside the channel.
Analyzing a large parameter space which encompasses particle properties and source position, mutual distance and alignment with respect to the mean flow, we were able to provide a thorough description of mixing in a complex setting such as anisotropic turbulence. 
The definition of an encounter probability to represent the likeliness that particles released from distinct sources mix is an effective tool that relies solely on a Lagrangian description of particle motion.

We have shown how the evolution of $E$ through time is a result of the competing interaction of processes that bring particles together and disperse them, acting on a range of flow scales. 
We demonstrated the highly anisotropic features of mixing near the wall, where mixing of streamwise-aligned sources is far larger than that of those aligned along the wall-normal direction. Indeed, the tendency of particles released very close to the wall to organize themselves into elongated streaks aligned with the direction of the mean flow favors this directional imbalance between mixing properties. Further away from the wall, the encounter probabilities tend to become independent of the direction of alignment of sources, as channel properties also tend to isotropy given the reduced influence of the solid boundary. Still, while isotropy is attained at lower Stokes number, this is not the case for particles with larger inertia where wall-normal mixing is hindered at the center of the channel. We also computed encounter properties for pairs of sources aligned along the spanwise direction, obtaining results that are quite similar to those obtained from $y$-aligned sources (although encounter rates near the wall are higher due to the absence of mean shear).

Effects of particle inertia are evident throughout the results reported in this work, although the properties of the encounter probability do not have a straightforward dependence on the Stokes number. More precisely, the effects of varying particle inertia are different depending on the position of source pairs inside the channel and on their alignment. Indeed, the role played by inertial clustering depends heavily on the ratio between the particle response time and the smallest relevant timescale of the flow $\tau_{\eta}$, \textit{i.e.} $St_K$. The latter depends on the distance from the wall and in particular increases away from it, ranging from values of around $\tau_{\eta}^+ = 2$ near the wall to around 30 at the center of the channel. As the effects of inertial clustering are the most intense when $St_K = St^+/\tau_{\eta}^+$ is of order unity, particles experience clustering not only depending on their inertia, but also on their position inside the channel. Indeed, a particle experiencing strong inertial clustering near the center of the channel could move towards the wall and start reacting weakly to turbulent fluctuations as its (local) Stokes number $St_K$ increases.
The encounter rate and the medium- to long-term evolution of the mixing, which are the focus of this work, appear to be affected by the formation of small-scale clusters, which happens on a short timescale \citep{liu2020jfm}. 

Near the wall, peak encounter rates of streamwise-aligned sources are achieved in such short times (compared to the integral velocity timescale of particles) that no differences due to the Stokes number arise. On the other hand, for wall-normal-aligned sources, values of $E_{peak}$ are slightly higher at higher Stokes numbers, albeit still very small. Moving farther away from the wall, particles with $St^+ = 5$ first and $St^+ = 25$ afterwards experience the effects of inertial clustering as their typical timescale matches locally that of the flow, possibly resulting in the increases of the peak encounter rate seen near $y^+ = 100$ at $St^+ = 5$ and $y^+ = 900$ at $St^+ = 25$.

Accordingly, two different processes emerge as the cause of the observed trends of the encounter rate. 
The highly anisotropic features of near-wall turbulence, which result in elongated and highly persistent streaks of particles, determine strong mixing of streamwise-aligned sources (and weak mixing of wall-normal-aligned sources) on timescales so short that inertial effects caused by the matching between the particle response time, $\tau_p$, and the smallest timescales of the flow are negligible. Here it is possible that the relevant timescale of the encounter rate evolution is linked to the Lagrangian integral timescale, highlighting the importance of the initial condition that determines the very-short term evolution of particle motion.
On the other hand, away from the wall the properties of the encounter rate are clearly influenced by inertial clustering, with encounter rates increasing where the particle timescale locally matches that of the flow.
Therefore, both the Kolmogorov timescale of the flow and the Lagrangian integral timescale of particles influence the encounter rate and, by consequence, mixing.

In order to test the sensitivity of results on the Reynolds number, we simulated mixing for $\ret = 590$ (see \ref{app:a}). We observed that the measures presented previously have a reduced sensitivity on the Reynolds number (at least in the range of $\ret$ considered), except the fact that the evolution of the encounter probability takes place on a slightly longer timescale. Furthermore, we tested our choice of the threshold value $R$ showing that it is appropriate to capture the organization of particles into clusters due to their inertia (see \ref{app:b}). Indeed, at larger values of $R$ most of the differences between particles at different Stokes numbers disappear. Conversely, the variation of the threshold value is a simple way to investigate the typical scale at which inertial clustering effects are relevant for mixing processes.

Differently from methods that investigate mixing of particles by, for example, binning particles into fixed partitions of the domain, the method used in this work is purely Lagrangian, that is only particle trajectories and their mutual distances are analyzed. This is advantageous as it eliminates the need (and the arbitrariness) to define bins and rather uses only particle mutual distance as a criterion \citep{nguyen2018aj}. 
Further developments may be conducted by relating particle distributions with the underlying local properties of the turbulent flow field, in order to provide a deeper characterization of the interplay of near-wall coherent structures, inertial clustering phenomena and velocity fluctuations observed in this work.

Overall, Lagrangian based methods emerge naturally when mixing and dispersion problems are studied, and are especially convenient in complex settings such as those represented by turbulent flows.
A deeper understanding of basic mixing processes achieved through high resolution simulation is instrumental in developing better simplified models of mixing. This is especially true given the relevance of processes involving inertial particles, where the formation of particle clusters may greatly influence processes such as turbulent combustion.

\section*{Acknowledgements}
This work was sponsored by NWO Exacte en Natuurwetenschappen (Physical Sciences) for the use of supercomputer facilities, with financial support from the Netherlands Organization for Scientific Research, NWO.

\appendix
\section{Effects of the Reynolds number}
\label{app:a}
In order to test the dependency of the results presented in this work on the Reynolds number $\ret$, we performed the same analysis at $\ret = 590$. To do so, we performed a direct numerical simulation of the same channel flow geometry, integrating the trajectories of particles at different Stokes numbers using the same release configuration as before. The grid of the DNS was left unchanged, while the time-step was $\Delta t^+ = 0.07375$.

\begin{figure}
\centering
\includegraphics[width = \textwidth]{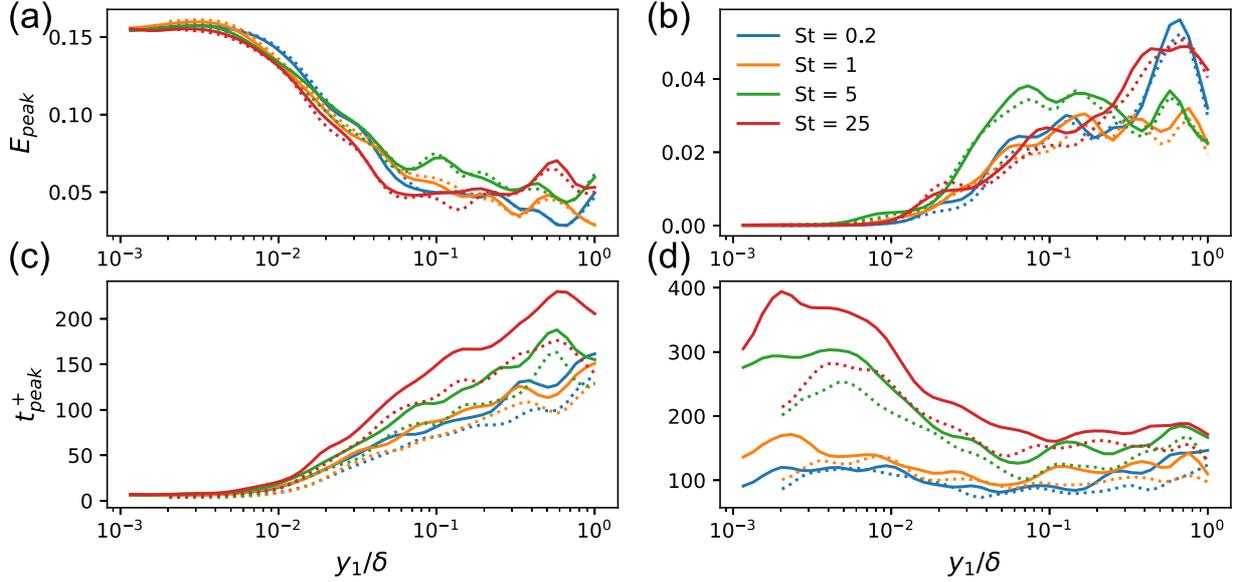}
\caption{Streamwise (a) and wall-normal (b) peak encounter probability at $\ret = 950$ (solid lines) and $\ret = 590$ (dotted lines). $d^+ = 4$ in all cases. Peak time of the streamwise (c) and wall-normal (d) encounter probability at $\ret = 950$ (solid lines) and $\ret = 590$ (dotted lines). \label{fig:a1}}
\end{figure}

Figure \ref{fig:a1} shows the effects of the lower Reynolds number on the encounter probability, and especially on the peak value and on the time of the peak $t_{peak}$. In all plots we normalize the wall-normal coordinate using the value of $\delta$ (the channel half-height), so that both configurations are compared in outer units. 
Figures \ref{fig:a1}(a) and \ref{fig:a1}(b) show the streamwise and wall-normal peak encounter probability, respectively. As can be noted, the Reynolds number has little to no influence on the value of $E_{peak}$ along both directions.
On the other hand, the peak (see figures \ref{fig:a1}(c) and \ref{fig:a1}(d)) occurs at an earlier time (in wall units) at the lower Reynolds number, both in the case of streamwise mixing for sources located away from the wall and everywhere in the channel in the case of wall-normal mixing. If outer units are used for time (i.e. $t = t^+ u_{\tau}/\nu$), no Reynolds independence is observed either, and $t_{peak}$ occurs at a later time instead at $\ret = 590$. Therefore there is no clear scaling of peak times with respect to the Reynolds number in the range here considered.

\section{Direct numerical simulation validation}
\label{app:c}

\begin{figure}
\centering
\includegraphics[width = \textwidth]{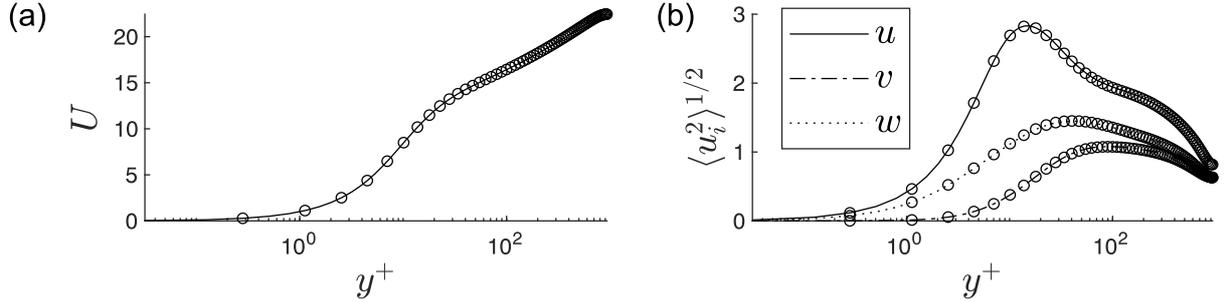}
\caption{Mean (a) and root mean square (b) velocity profiles in wall units of the Eulerian phase of the channel flow simulation. Solid lines: present simulation; dots: data from \cite{hoyas08}.\label{fig:a3}}
\end{figure}

Mean and root mean square velocities from the channel flow were computed for the streamwise, wall-normal and spanwise components and are shown in figure \ref{fig:a3}. Results were obtained by computing statistical quantities over several flow-through times. The present results, shown with solid lines in figure \ref{fig:a3}, were compared with those from \cite{hoyas08} (dots) and found in good agreement.

\section{Effects of the threshold distance R}
\label{app:b}
\begin{figure}
\centering
\includegraphics[width = \textwidth]{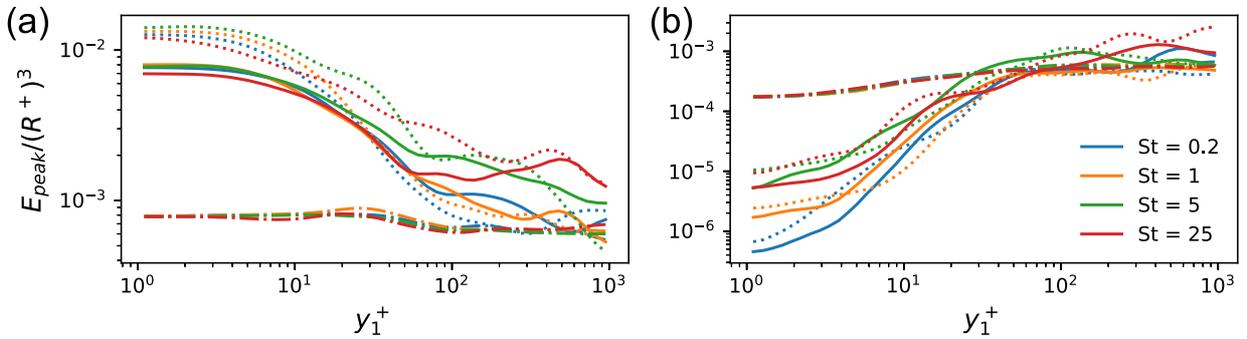}
\caption{Streamwise (a) and wall-normal (b) peak encounter probability at varying Stokes number and $d^+ = 20$. In all panels, solid lines indicate $R^+ = 2$, dotted lines $R^+ = 1$, and dash-dotted lines $R^+ = 8$ (a logarithmic scale is used on the vertical axis).\label{fig:a2}}
\end{figure}

As stated before, the choice of the value of the threshold $R$ is instrumental in describing the key interactions between particles that lead to their mixing, especially in relation to the length scales typical of inertial clustering of particles with $St>0$. We evaluated, using the same setup as before, the influence of the value of $R$ by computing the peak encounter probability properties with $R^+ = 1$ and $R^+ = 8$. The results are shown, along with those obtained using $R^+ = 2$ as before, in figure \ref{fig:a2}.
Figures \ref{fig:a2}(a) and \ref{fig:a2}(b) show the peak encounter probability normalized by $\left(R^+\right)^3$. In all cases, $d^+$ has been chosen equal to 20 in order to avoid that sources have a nonzero encounter probability at the time of release. Indeed, it can \textit{a priori} be assumed that the encounter rate is proportional to the volume of the sphere of influence surrounding each particle. The larger the volume of the sphere, the larger the number of interactions that will be experienced by each particle. In reality, this is shown to be approximately true only in some cases. In the case of streamwise mixing, the independency of $E_{peak}/\left(R^+\right)^3$ from $R$ is only attained in a weak manner near the center of the channel (where differences between mixing at different $St$ are minimal and appear amplified by the logarithmic scale), while near the wall and up to about $y^+ = 200$ the peak encounter probability grows more slowly than $R^3$ when $R$ is increased. On the other hand, the collapse of the peak encounter probabilities with different values of the threshold distance is stronger in the case of wall-normal mixing and is attained for source pairs at a lower $y^+$ coordinate. Furthermore, the peak encounter probability grows faster than $R^3$ when $R$ is increased, in contrast to what happens for streamwise mixing.
The intense anisotropy of turbulence in the near-wall region and the presence of the wall highly skew mixing processes in certain directions. This may cause in turn a non-homogeneous growth of the encounter probability with respect to the threshold distance $R$, which grows faster than expected along $\mathbf{\hat{e}}_y$ and slower along $\mathbf{\hat{e}}_x$.

Furthermore, we note that at $R^+ = 8$ the differences between peak encounter probabilities at different Stokes numbers are vanishing, which is a sign that the threshold is larger than the length scales typical of inertial clustering and thus differences due to inertia do not emerge anymore. Indeed, this indicates that mixing processes involving particles with larger inertia are not always influenced by inertial clustering, only those happening at sufficiently small length scales. Overall, similar trends are observed across this range of threshold values, although the sensitivity to inertial clustering decreases with increasing values of $R$.

\bibliographystyle{elsarticle-harv} 
\bibliography{biblio}

\end{document}